%% file: pap97.tex
\documentstyle[12pt,rotating,epsfig]{cernart}
\renewcommand{\marginparwidth}{0.6cm}
\renewcommand{\marginparsep}{0.2cm}
\makeatletter
\def\thebibliography#1{\section*{\refname\@mkboth
 {\uppercase{\refname}}{\uppercase{\refname}}}\list
 {\@biblabel{\arabic{enumi}}}{\settowidth\labelwidth{\@biblabel{#1}}%
 \leftmargin\labelwidth
 \advance\leftmargin\labelsep
 \usecounter{enumi}
 \def\theenumi{\arabic{enumi}}}%
 \def\newblock{\hskip .11em plus.33em minus.07em}%
 \sloppy\clubpenalty4000\widowpenalty4000
 \sfcode`\.=1000\relax}
\makeatother
\def\Aref#1{$^{\rm #1}$} 
\def\AAref#1#2{$^{\rm #1,#2}$} 
\def\IAref#1#2{$^{\Inst{#1},\rm #2}$} 
\def\IIref#1#2{$^{\Inst{#1},\Inst{#2}}$} 

\def\IIAref#1#2#3{$^{\Inst{#1},\Inst{#2},\rm #3}$}
\def\Iref#1{$^{\Inst{#1}}$} 
\begin {document}

\date {Nov 11 1997}
\vglue2cm
\title{POLARISED QUARK DISTRIBUTIONS IN THE NUCLEON FROM SEMI-INCLUSIVE SPIN ASYMMETRIES}
\vglue2cm
\collaboration {The Spin Muon Collaboration (SMC)}
\vspace{1cm}
\vglue4cm
\begin{abstract}
We present a measurement of semi-inclusive spin asymmetries for positively 
and negatively charged hadrons  from deep inelastic scattering 
of polarised muons on polarised protons and deuterons in the range 
$0.003<x<0.7$ and $Q^2>$1~GeV$^2$. 
Compared to our previous publication on this subject,
with the new data the statistical errors have been reduced by nearly a factor of two.
From these asymmetries and our
inclusive spin asymmetries we determine the
polarised quark distributions of valence quarks and
non-strange sea quarks at $Q^2$=10~GeV$^2$.
The polarised $u$ valence quark distribution, $\Delta u_v(x)$, is positive and the polarisation
increases with $x$.
The polarised $d$ valence quark distribution, $\Delta d_v(x)$, is negative
and the non-strange sea distribution, $\Delta \bar q(x)$, is consistent with zero over the measured
range of $x$.
We find for the first moments 
$\int_0^1 \Delta u_v(x) {\rm d}x = 0.77 \pm 0.10 \pm 0.08$,
$\int_0^1 \Delta d_v(x) {\rm d}x = -0.52 \pm 0.14 \pm 0.09$ 
and $\int_0^1 \Delta \bar q(x) {\rm d}x= 0.01 \pm 0.04 \pm 0.03$,
where we assumed $\Delta \bar u(x) = \Delta \bar d(x)$.
We also determine for the first time the second moments of the valence
distributions $\int_0^1 x \Delta q_v(x) {\rm d}x$.
\end{abstract}
\vspace {1cm}
\submitted {To be submitted to Physics Letters B}

\newpage
\input{authors}
\pagestyle{plain} 
   Measurements of spin asymmetries in polarised
deep inelastic scattering provide 
information about the spin structure of the nucleon.
In particular detailed information can be obtained from polarised semi-inclusive deep
inelastic scattering, $\vec \mu + \vec N \rightarrow \mu + X + h$, 
where in addition to the scattered lepton hadrons ($h$)
are also detected.
In this paper we present new results on semi-inclusive asymmetries from SMC data including those 
published in ~\cite{smc_si}.
We analyse these asymmetries together with the inclusive asymmetries
in the framework of the quark parton model (QPM) and determine
the polarised quark distributions of the valence quarks and of the non-strange
sea quarks.  
This is only possible due to the semi-inclusive data.

Our experimental setup at the CERN muon beam consists 
of three major components: a polarised target, a magnetic spectrometer and 
a muon beam polarimeter. 
A detailed description of the experiment and the analysis of
the inclusive data can be found in ref.~\cite{smc_pprep,smc_plb396}. 
Positive muons of a nominal energy of 100 and 190 GeV were used.
Most of the data were taken at 190~GeV.
The muon beam polarisation $P_B$ was determined from measured spin
asymmetries in muon-electron scattering and for the 190 GeV data in addition  
from the energy spectrum of positrons from muon decays.
It was found to be $-0.795 \pm 0.019$
\footnote{This value results from a recent
analysis, to be published.}     ($-0.81\pm 0.03$) for an average beam energy of 187.4~GeV (99.4~GeV).
The target consisted of two cells filled with butanol, deuterated butanol or
ammonia.
The two cells were polarised in opposite directions by 
dynamic nuclear polarisation. 
The average polarisations $P_T$ were approximately $0.90$ for protons 
and $0.50$ for deuterons with a relative error $\Delta P_T/P_T$ of $3\%$.
The determination of semi-inclusive asymmetries requires the separation of hadrons
from electrons which originate mainly from photon conversions. For this purpose
we used a calorimeter \cite{h2_lit} which consists of an electromagnetic and a
hadronic part.
The electromagnetic part amounts to 20 radiation lengths which is 
sufficient to contain electromagnetic showers.
The total thickness of the calorimeter is 5.5 nuclear interaction lengths.
For each shower the ratio of the energy deposited in 
the electromagnetic part to the total deposited energy is calculated. 
Electrons are eliminated by requiring this ratio to be smaller than $0.8$.
There was no hadron identification, only the charge of the hadron is known.

The asymmetries of the spin-dependent 
virtual photon absorption cross sections 
for production of positive (negative) hadrons 
are defined as
\begin{equation}\label{eq01}
A_1^{+(-)} = \frac{\sigma^{+(-)}_{\uparrow \downarrow} - \sigma^{+(-)}_{\uparrow \uparrow}}
{\sigma^{+(-)}_{\uparrow \downarrow} + \sigma^{+(-)}_{\uparrow \uparrow}} \, ,
\end{equation}
where the indices $\uparrow \downarrow$ and $\uparrow \uparrow$ refer to the relative
orientation of the nucleus (proton or deuteron)  and  photon spins.
Contributions from the asymmetry $A_2$ are neglected.
The cross sections $\sigma^{+(-)}$ refer to the number of
produced hadrons, i.e. all hadrons detected in one event
are counted.
The extraction of the semi-inclusive asymmetries from the counting rates is described
in \cite{smc_si}.

The counting rate asymmetry $A^{exp}$ is related to the virtual
photon asymmetry $A_1$ by:
\begin{equation}
A_1 = \frac{1}{P_B P_T D f} A^{exp} \, .
\end{equation}
The depolarisation factor $D$ depends on the event kinematics and on the ratio 
$R$ of longitudinal to transverse  virtual photon 
cross sections \cite{smc_pprep}.
The dilution factor $f$ accounts for the presence of unpolarisable
nuclei in addition to the protons or deuterons in the target.
It can be expressed in terms of the numbers $n_A$ of nuclei with mass 
number $A$ and the corresponding total spin-independent cross sections 
$\sigma_A^{tot}$.
Taking into account radiative effects on the nucleon \cite{smc_pprep,terad}
the effective dilution factor is defined as
\begin{equation}\label{dilf}
f=\frac{\sigma_{p,d}^{1\gamma}}{\sigma_{p,d}^{tot}} \,
\frac{n_{p,d} \sigma_{p,d}^{tot}}{\sum_A n_A \sigma^{tot}_{A}} \, .
\end{equation}
Here $p,d$ stands for proton or deuteron.

The evaluation of the dilution factor for inclusive events, $f^i$, is
described in ref.~\cite{smc_pprep}.
The main contributions to the total inclusive cross section are 
the one photon exchange with vertex and vacuum polarisation corrections,
and the inelastic, elastic and quasielastic bremsstrahlung processes.
In the elastic and quasielastic bremsstrahlung processes no hadrons are produced
and therefore these processes do not contribute to the
total semi-inclusive cross section.
In the analysis hadrons are selected if their energy is above a certain fraction, $z$,
of energy transfer $\nu$.  
This reduces the energy available for a possible bremsstrahlung photon accompanying
an inelastic event, relative to the inclusive case.
The contribution from inelastic bremsstrahlung is then reduced.
This reduction was estimated to be small and is included only in the systematic error
of the dilution factor.
We compute the dilution factor for the semi-inclusive events, $f^{si}$,
without the elastic and quasielastic contributions in the cross section.
In the inclusive case, at low $x$ there is a large contribution of elastic bremsstrahlung from
high $Z$ nuclei in the denominator of eq.~(\ref{dilf}).
It is not present
 in the semi-inclusive case and therefore $f^{si}$ is 30\% larger than $f^i$
in this region,
whereas the difference
is very small for $x>0.1$.
Using the QPM we compute from $f^{si}$  
the dilution factors for positive, $f^{+}$, and negative, $f^{-}$, hadrons~\cite{papav,joerg}.
For the polarised deuterated butanol target $f^+$ and $f^-$ are equal to $f^{si}$.
For the butanol and ammonia targets, where protons are polarised,
the ratio $f^+/f^{si} \approx 1.07$ and $f^-/f^{si} \approx 0.88$  
at $x=0.5$ and both ratios are close to 1 at low $x$.
The ratios differ from unity for these  targets because 
more positive hadrons are produced on protons 
at large $x$ as compared to the isoscalar unpolarised target nuclei.
The correctness of the above procedure for the evaluation of the dilution factor 
was verified using a Monte Carlo in which radiative processes and hadron production were simulated.

Polarised radiative corrections are applied to the asymmetries as described in 
ref.~\cite{smc_pprep,polrad}. In this procedure, as for the evaluation of the dilution factor, contributions of
processes where no hadron is produced are omitted.

In order to interpret the results in the QPM a cut of $Q^2>1$ GeV$^2$
is applied. The current fragmentation region is selected by a cut on $z=E_{had}/\nu>0.2$.
After cuts on $\nu$, $y$, the 
energy and the angle of the scattered muon as in ref.~\cite{smc_pprep},
we obtain $32 \times 10^6$
events. After applying the $z$ cut we obtain $5 \times 10^6$ positive and 
$4 \times 10^6$ negative hadrons. 
All of the data are in the 
deep inelastic region, $W>3$~GeV,
and cover the range $0.003<x<0.7$ with 
an average $Q^2 \approx 10$~ GeV$^2$. 

\begin{figure}[H]
\begin{center}
\mbox{\epsfig{file=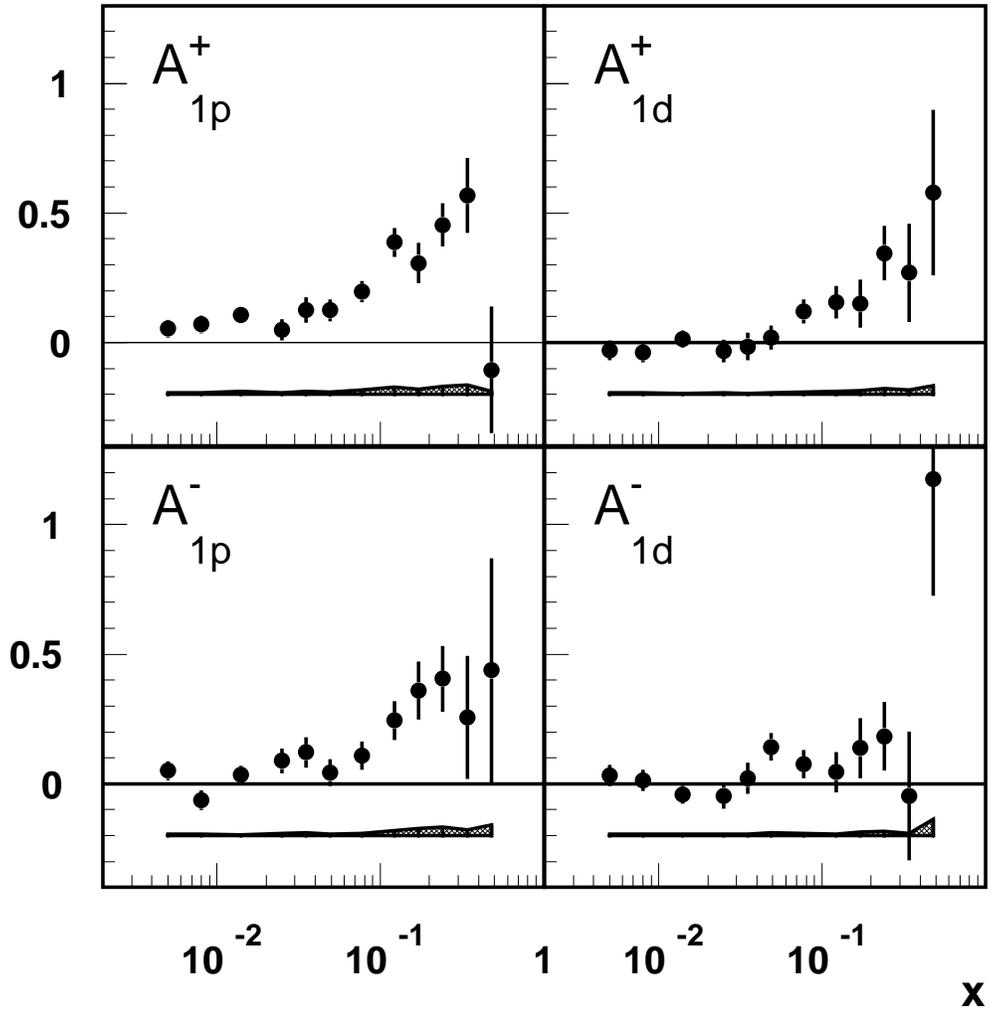,width=\textwidth}}
\end{center}
\caption{\em Semi-inclusive spin asymmetries 
for the proton and the deuteron as a function of $x$ for positively and
negatively charged hadrons.
The error bars are statistical and the shaded areas represent the systematic 
uncertainties. \label{asym_fig}}
\end{figure}

The asymmetries measured in different periods of data taking
are compatible with each other, and thus all SMC data are combined.
The semi-inclusive asymmetries $A_{1p,d}^{+,-}$ for positive and negative 
hadrons from the deuteron and the proton  
are presented in Fig.~\ref{asym_fig} and Tab.~\ref{asym_tab}.
The correlations between the asymmetries
are listed in Tab.~\ref{cor_tab}.
The main contributions to systematic errors are due to the 
uncertainties of the target
and beam polarisations, to the variation
in time of the spectrometer acceptance and to the uncertainty of the dilution factor.
Negligible contributions arise from 
secondary interactions of hadrons in the target, radiative corrections and
differences in acceptance for pions, 
kaons and protons due to different absorptions in the target and different 
angular distributions.

The inclusive asymmetry $A_{1p}$ can be expressed in terms of polarised 
and unpolarised structure functions:
\begin{equation}\label{a1incl1}
A_{1 p}(x,Q^2) = \frac{g_1^p(x,Q^2)}{F_1^p(x,Q^2)} =
\frac{g_1^p(x,Q^2)}{F_2^p(x,Q^2)} \, 2x \, \left[ 1+ R(x,Q^2) \right] \, .
\end{equation}
In the QPM the structure functions can be written in terms of polarised
quark distributions $\Delta q = q^{\uparrow} - q^{\downarrow}$ 
and unpolarised quark distributions
$q= q^{\uparrow} + q^{\downarrow}$, 
where $q^{\uparrow}(q^{\downarrow})$ is the
distribution of quarks of flavour $q$ and spin
parallel (antiparallel) to the nucleon spin.
The published leading order parametrisations of the unpolarised quark
distributions were obtained from experimental values of $F_2^p$ using the relation
$F_2^p(x,Q^2)=x \sum_q e^2_q \left[ q(x,Q^2) + \bar q(x,Q^2) \right]$.
The values of $F_2^p$ were extracted from cross sections assuming non-zero values of $R$.
Therefore the expression for $A_{1p}$ becomes:
\begin{equation}\label{a1incl2}
A_{1 p}(x,Q^2) =\frac{\sum_q e_q^2\, \left[ \Delta q(x,Q^2) + \Delta \bar q(x,Q^2) \right]}
{\sum_q e_q^2\, \left[ q(x,Q^2) + \bar q(x,Q^2) \right] } \, \left[ 1+R(x,Q^2) \right] \, ,
\end{equation}
where $e_q$ is the fractional charge of the quark $q$ with 
$q=u,d,s$.
For $x < 0.12$ we use a parametrisation of $R$
measured by the NMC~\cite{rnmc} and for higher $x$ we use 
the SLAC parametrisation ~\cite{rslac}.
At our average $Q^2$ of 10~GeV$^2$ the ratio $R$ varies from
0.15 at $x=0.005$ to 0.07 at $x=0.5$.

For semi-inclusive processes the asymmetries 
depend in addition on the
fragmentation functions $D_q^h(z,Q^2)$, which represent
the probability that a quark $q$ fragments into a hadron $h$.
Using the quantity  $D_q^h(Q^2)$ defined as
 $D_q^h(Q^2)=\int_{0.2}^1 {\rm d}z\,D_q^h(z,Q^2)$ we have:
\begin{equation}\label{a1si}
A_{1p}^{+(-)}(x,Q^2)=\frac{\sum_{q,h} e_q^2\,
       \left[ \Delta q(x,Q^2)\,D_q^h(Q^2) + \Delta \bar q(x,Q^2)\,D_{\bar q}^h(Q^2) \right]}
 {\sum_{q,h} e_q^2\, \left[ q(x,Q^2)\,D_q^h(Q^2) 
+ \bar q(x,Q^2)\,D_{\bar q}^h(Q^2) \right]} \, 
\left[ 1+R(x,Q^2) \right] \, ,
\end{equation}
where the summation is over $\pi^+,K^+$ and $p$ for positive hadrons
and over $\pi^-,K^-$ and $\bar p$ for negative hadrons.
We assumed the same $R$  for 
positive and negative hadrons, and used the same $R$ as for the inclusive case.
It was checked that our acceptance depends only weakly on $z$
 in the region covered by the integral of the fragmentation functions.
Since the fragmentation process is parity conserving and the polarisation 
of hadrons is not observed, the fragmentation functions do not depend on the
quark helicity ($D_{q^\uparrow}=D_{q^\downarrow}$). 
In general $D_q^h \ne D_{\bar q}^h$, so 
the measurement of semi-inclusive asymmetries allows a separation
of $\Delta q$ and $\Delta \bar q$, whereas 
only the sum, $(\Delta q + \Delta \bar q)$, can be determined
from the inclusive asymmetries. 

Using isospin symmetry, similar expressions are obtained for the deuteron asymmetries.
From the data on proton and deuteron we can separate spin contributions from
$u$ and $d$ quarks.
The spin dependent deuteron cross section is considered to be the sum of the 
proton and the neutron cross sections with a correction to account for
the $D$-state probability $\omega_D = 0.05 \pm 0.01$
of the deuteron, as in ref.~\cite{smc_plb396}.
We use the same $R$ for the proton and the deuteron.

With eqs.~(\ref{a1incl2}) and (\ref{a1si}) 
and the corresponding relations for the deuteron
our measured inclusive
and semi-inclusive asymmetries can be used to evaluate
the polarised quark distributions.
The polarised valence quark distributions are defined as
$\Delta u_v(x) = \Delta u(x) - \Delta \bar u(x)$ and
$\Delta d_v(x) = \Delta d(x) - \Delta \bar d(x)$.
To reduce the number of unknowns we assume a $SU(3)_f$ symmetric sea which we denote
by $\Delta \bar q(x)$:
\begin{equation}\label{su3sea}
\Delta \bar q(x) = \Delta \bar u(x) = \Delta \bar d(x) = \Delta s(x) = \Delta \bar s(x) \, .
\end{equation}
The sensitivity of our results to this assumption will be discussed below. 
In particular the assumption involving $\Delta s(x)$ and $\Delta \bar s(x)$
has a negligible influence on our results,
because our hadron sample contains mainly pions, thus $\Delta \bar q(x)$  provides
information mainly about non strange sea quarks.

The six measured
asymmetries are linear combinations of three unknowns:
$\Delta u_v$, $\Delta d_v$ and $\Delta \bar q$.
Six equations for the asymmetries can thus be written in matrix form:
\begin{eqnarray}\label{lgs}
&\vec A = {\cal B} {\Delta \vec q}  &
\end{eqnarray}
where $\vec A =  \Big( A_{1p},  A_{1p}^+,  A_{1p}^-, A_{1d}, A_{1d}^+,A_{1d}^- \Big)$
and $\Delta \vec q = \Big( \Delta u_v,\Delta d_v,\Delta \bar q \Big)$.
The elements of the matrix $\cal B$ 
are determined from eqs.~(\ref{a1incl2}) and (\ref{a1si}).
They depend on the unpolarised
quark distributions $q$, the fragmentation functions $D_q^h$,
the ratio $R$, and $\omega_D$.

The asymmetries were measured at average $Q^2$ values in each $x$ bin
varying from 1.3~GeV$^2$ at our lowest $x$ to 60~GeV$^2$ at $x=0.5$.
No significant $Q^2$ dependence is observed in the inclusive data~\cite{smc_plb396,prot96},
therefore we assume the asymmetries at measured $Q^2$ to be equal  
to those at $Q^2=10~$GeV$^2$.
In eq.~(\ref{lgs}) we use parametrisations of the unpolarised quark 
distributions~\cite{grv}, of the ratio $R$,
and of the fragmentation functions ~\cite{emcff}, all
at  $Q^2=10$~GeV$^2$.
For the unmeasured fragmentation functions additional 
assumptions are made, as discussed below.

\begin{figure}[H]
\begin{center}
\mbox{\epsfig{file=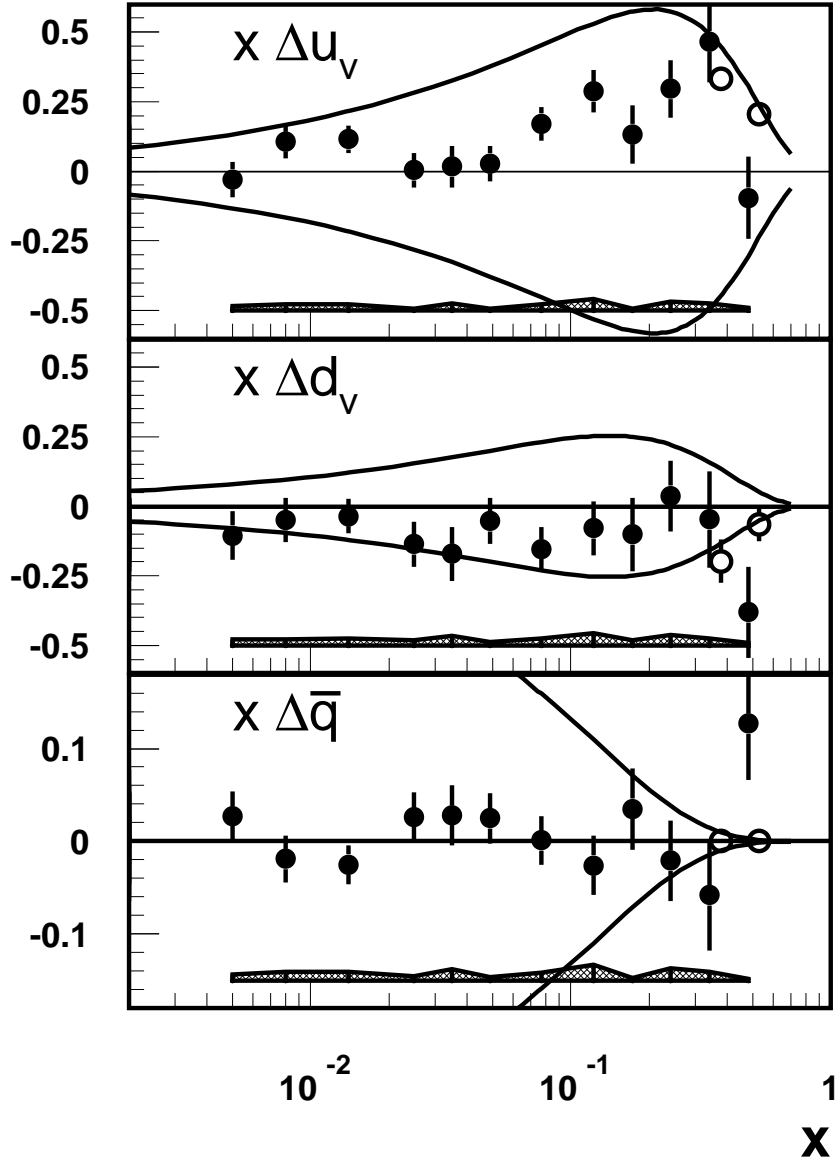,width=\textwidth}}
\end{center}
\caption{\em The polarised quark distributions
$x \Delta {u}_v(x)$, $x \Delta {d}_v(x)$ and $x \Delta \bar{q}(x)$ 
obtained with the assumption
$\Delta \bar{u}(x)=\Delta \bar{d}(x)$.
The open circles are obtained when the sea polarisation 
is set to zero while the closed circles are obtained without
this assumption.
The error bars are statistical and the shaded areas represent the systematic 
uncertainty. The curves correspond to the upper and the lower limits 
$\pm x q(x)$
from the unpolarised quark distributions~\cite{grv} evaluated at $Q^2=10$~GeV$^2$.
In the bottom plot the curves are $\pm x(\bar u(x) + \bar d(x))/2$.
\label{dq}}
\end{figure}

In order to determine the leading order polarised quark distributions
$\Delta u_v(x)$, $\Delta d_v(x)$ and $\Delta \bar q(x)$ 
we solve eq.~(\ref{lgs}) in every $x$-bin independently by 
minimising the quantity
\begin{equation}
\chi^2 = (\vec A - {\cal B} \vec{\Delta q})^T \,
(\mbox{Cov}_{A})^{-1} \, (\vec A - {\cal B} \vec{\Delta q}) \, ,
\end{equation}
where $\mbox{Cov}_{A}$ is the covariance matrix of the asymmetries
(cf. Tab.~\ref{cor_tab}).
The results are shown in Fig.~\ref{dq} and Tab.~\ref{dq_tab}.
We see in the figure that
the statistical error on $\Delta \bar{q}(x)$ in the region $x>0.3$ 
is larger than $\bar{q}(x)$ which is an upper limit for $|\Delta \bar{q}(x)|$.
In this region, in order to reduce the statistical error on
$\Delta u_v(x)$ and $\Delta d_v(x)$, we set $\Delta \bar{q}(x)=0$ before 
solving the system of equations. 
The results are shown as open circles in Fig.~2.
Uncertainties due to this assumption were included in the systematic error.

\begin{figure}[H]
\begin{center}
\mbox{\epsfig{file=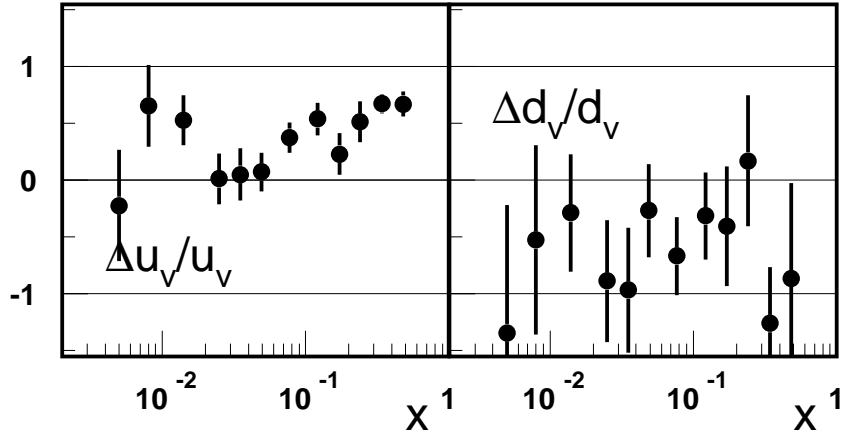,width=\textwidth}}
\end{center}
\caption{\em The polarisation $\Delta q_v(x)/q_v(x)$ of the valence
quarks. Only the statistical errors are given.
\label{pol}}
\end{figure}

We observe that $\Delta u_v(x)$ is positive while $\Delta d_v(x)$ is negative.
The polarised sea quark distribution $\Delta \bar q(x)$ 
is compatible with zero over the full range of $x$.
Fig.~\ref{pol} shows the polarisation $\Delta q_v(x)/ q_v(x)$ of
the valence quarks. The average polarisation  
is approximately $50\%$ for the $u_v$ quarks and $-50\%$ for the 
$d_v$ quarks. For the $u_v$ quarks we observe a lower
average polarisation of $0.18\pm0.10$ in the interval $0.003<x<0.06$ (first six bins)
and a higher average polarisation of $0.57\pm0.05$ in the interval 
$0.06<x<0.7$ (last six bins).

The uncertainties of these results are dominated by the 
statistical errors.
Contributions from different sources to the  
systematic errors on the integrals $\int_{0.003}^{0.7} \Delta q(x) {\rm d}x$
are listed in Tab.~\ref{syserr} and the values of the integrals in Tab.~\ref{int}.
The largest contribution is related  to the assumptions made for the unmeasured 
fragmentation functions of $s$ quarks.
Strange quark fragmentation functions to pions were assumed to be equal
to the measured unfavoured fragmentation function:
$D_{s,\bar s}^{\pi^+,\pi^-} = D_u^{\pi^-}$.
For the favoured fragmentation functions of $s$ quarks to kaons 
we assumed
$D_{s}^{K^-} = D_{\bar s}^{K^+} = D_d^{\pi^-}$.
This is motivated by the fact that in both
fragmentation processes, $s \rightarrow K^-$ and $d \rightarrow \pi^-$,
a $u \bar u$-pair has to be produced to create the hadron.
Values of the unmeasured fragmentation functions obtained with the above assumptions 
were varied by a factor 0.5 to 1.5 
and the difference in the resulting polarised quark distributions was included in the systematic error.
The uncertainty due to the unpolarised quark distributions
was taken to be the difference resulting from 
two different parametrisations, GRV-94 (LO)~\cite{grv} and CTEQ 3L (LO)~\cite{cteq}.
The systematic error due to the assumption $\Delta \bar q(x)=0$
in the last two $x$ bins was calculated by setting 
$\Delta \bar q(x) = \pm \bar q(x)$ in these bins.

\begin{figure}[H]
\begin{center}
\mbox{\epsfig{file=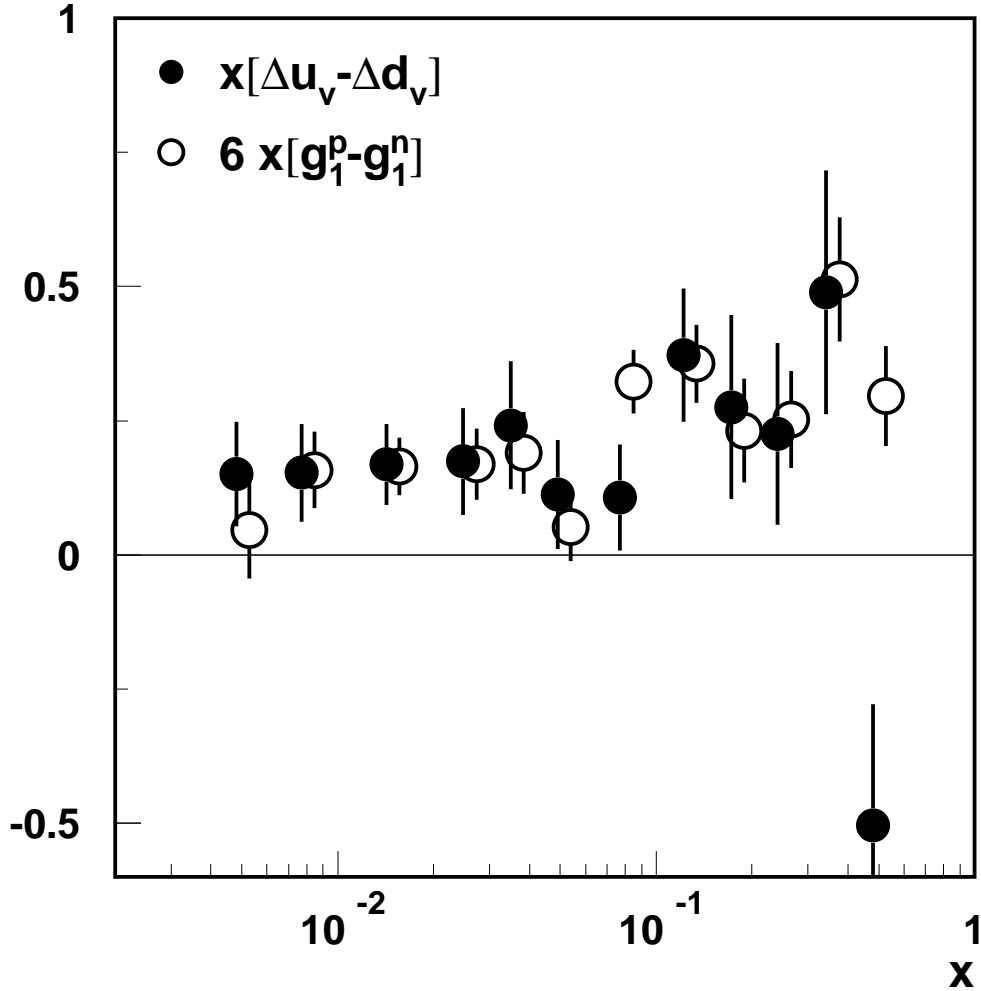,width=\textwidth}}
\end{center}
\caption{\em The difference between the spin-dependent 
structure functions of the proton and the neutron,
$6\,x[g_1^p(x)-g_1^n(x)]$, (open circles),
and the difference between polarised {\it up} and {\it down} valence quark 
distributions $x[\Delta u_v(x)-\Delta d_v(x)]$,
as determined with the assumption $\Delta \bar u(x) =\Delta \bar d(x)$ 
from SMC semi-inclusive asymmetries only
(closed circles). 
Errors are statistical.
\label{g1pg1n}}
\end{figure}

In the QPM, with the assumption $\Delta \bar{u}(x)=\Delta \bar{d}(x)$,
the quantities $6[g_1^p(x)-g_1^n(x)]$ and
$\Delta u_v(x)-\Delta d_v(x)$ are equal. 
The former is obtained from inclusive data only while
the latter can be extracted using only the four semi-inclusive 
asymmetries in eq.~(\ref{lgs}).
The two quantities are generally in good agreement, as seen in Fig.\ref{g1pg1n}, 
showing the consistency of the present
analysis.
The discrepancy in the last bin is mainly due to the low
value of the asymmetry $A_{1 \, p}^+$ in this bin.
This is also reflected in Tab.~\ref{dq_tab} by  
the low $\chi^2$ probability for this bin.
It has been checked that this low value could not be 
caused by a false asymmetry due to a change in the spectrometer
acceptance. 
With the constraint $\Delta \bar q(x)=0$ only two
unknowns, $\Delta u_v(x)$ and $\Delta d_v(x)$, have to be
determined. This can be done using only the inclusive
asymmetries $A_{1p}$ and $A_{1d}$ which have a better
statistical accuracy.
For this reason adding the semi-inclusive asymmetries has very little impact in the
last two bins, i.e. the results for
$\Delta u_v(x)$ and $\Delta d_v(x)$ from $A_{1p}$ and $A_{1d}$ only
are almost identical with the values
shown in Tab.~\ref{dq_tab} where all six asymmetries are used.

In order to calculate first moments, $\Delta q=\int_0^1 \Delta q(x) {\rm d}x$,
of the polarised quark distributions,
we need extrapolations of $\Delta q(x)$ to the unmeasured regions of $x$.
Contributions from the unmeasured large $x$ region ($0.7<x<1$) are negligible
because their upper limits given by the unpolarised distributions are small there.
The low $x$ ($0<x<0.003$) extrapolation was done in several ways.
Two functional forms of the polarised quark distributions
were fitted to our semi-inclusive and inclusive asymmetries~\cite{joerg}.
We also used parametrisations of the polarised quark distributions
obtained from two different QCD analyses ~\cite{grsv,gst} of inclusive data.
For every quark distribution
we obtained four values for the integral $\int_0^{0.003} \Delta q(x)$  and estimated the mean
value and the error. 
For the valence quarks these values (see Tab.~\ref{int}) are much smaller 
than the upper limits given by the unpolarised distributions  
$|\int_0^{0.003} \Delta q_v(x) {\rm d}x| \le \int_0^{0.003} q_v(x) {\rm d}x$. 
This upper limit is $0.22 \, (0.17)$ 
for GRV-94,LO parametrisation of the $u_v$ ($d_v$) quarks.
There is no such constraint for the sea quarks.
The first moments of the polarised quark distributions are
\begin{center}
\begin{tabular}{lcr}
$\Delta u_v$ &=& $ 0.77 \pm 0.10 \pm 0.08 $,  \\
$\Delta d_v $&=& $-0.52 \pm 0.14 \pm 0.09 $,  \\
$\Delta \bar{q}$&=&$0.01 \pm 0.04 \pm 0.03$. \\ 
\end{tabular}
\end{center}
The statistical errors are reduced by almost a factor of two compared
to our previous publication~\cite{smc_si}.
The contributions to the moments from measured and unmeasured regions
are detailed in Tab.~\ref{int}.

In order to study the effect of the assumption
of a $SU(3)_f$ symmetric sea in eq.~(\ref{su3sea})
we replace it by only an isospin symmetric sea, 
$\Delta \bar u(x) = \Delta \bar d(x) = \Delta \bar q(x)$
and allow $\Delta s(x) = \Delta \bar s(x) = \eta \Delta \bar q(x)$,
varying $\eta$ in the range $0.25<\eta<1.5$.
The corresponding variations of the first moments are of the order of 0.01.
Similar variations are observed if we
use $\Delta s(x)$ as given in ref.~\cite{grsv},
where the integral is $\int_0^1 \Delta s(x) {\rm d}x= -0.06$,
a value consistent with the result
$\int_0^1 (\Delta s(x)+ \Delta \bar s(x)) {\rm d}x= -0.12\pm 0.04$
obtained from the inclusive analysis~\cite{prot96}.
This shows that the measurement is only sensitive to the polarisation of
the non-strange sea quarks. 

Releasing the constraint $\Delta \bar u(x)=\Delta \bar d(x)$
and setting $\Delta s(x) = \Delta \bar s(x) = (\Delta \bar u(x) + \Delta \bar d(x))/2$
leads to the results given in the lower part of Tab.~\ref{int}.
The results are consistent with those obtained with the
constraint $\Delta \bar u=\Delta \bar d$.
The statistical error on $\Delta u_v(x)$
practically does not change, whereas the error on $\Delta d_v(x)$
increases by a factor of two.

Polarised semi-inclusive deep inelastic scattering is presently a unique
tool to measure the polarised valence and sea quark distributions.
However, the corresponding first moments can be determined by other methods. 
With the assumption of a $SU(3)_f$ symmetric sea 
(eq.~(\ref{su3sea})) 
and $SU(3)_f$ symmetry for the weak decays in the baryon octet, the first moments
of the valence distributions can be obtained from the axial matrix
elements, $F$ and $D$, of baryon decays.
It should be noted that the results depend strongly on these 
two assumptions, whereas the results from our semi-inclusive analysis
are insensitive to any $SU(3)_f$ assumptions, as discussed above.
With the values quoted in Ref.~\cite{par_dat,cl93} one finds
$\Delta u_v = 2F = 0.92 \pm 0.02$ and
$\Delta d_v = F-D = -0.34 \pm 0.02$,
which is in good agreement with our results
of Tab.~\ref{int} obtained from deep inelastic scattering only.
Our result for $\Delta \bar q$ is consistent with the first moment
of the polarised strange quark distribution
obtained from the inclusive analysis~\cite{prot96} quoted above.

From our polarised quark distributions we find for the second moments:
\begin{displaymath}
\int_0^1 x \Delta u_v(x)  {\rm d}x =  0.155 \pm 0.017 \pm 0.010 \; \; \mbox{and} \; \; 
\int_0^1 x \Delta d_v(x)  {\rm d}x = -0.056 \pm 0.026 \pm 0.011 \, ,
\end{displaymath}
where the first error is statistical and the second is systematic.
Due to the additional factor $x$ the low $x$ extrapolations to the unmeasured
region give negligible contributions while the large $x$ contributions remain small.
A calculation of the second moments
in lattice QCD~\cite{lech} predicts
$\int_0^1 x \Delta u_v(x)  {\rm d}x = 0.189 \pm 0.008 $ and
$\int_0^1 x \Delta d_v(x)  {\rm d}x = -0.0455 \pm 0.0032$,
which is in good agreement with our measured values.

To summarise, we have measured  semi-inclusive spin asymmetries
for positively and negatively charged hadrons from polarised protons
and deuterons and we have determined 
the polarised quark distributions $\Delta u_v(x)$, 
$\Delta d_v(x)$ and $\Delta \bar{q}(x)$ of 
the valence and the non-strange sea quarks.
The polarisation of the $u_v$ quarks was found to be
positive and to increase with $x$.
The polarisation of the $d_v$ quarks was found to be negative
while the polarisation of the non-strange sea quarks is consistent with zero.
Our results for the first moments of the polarised valence quark distributions are
in good agreement with results obtained from
the axial matrix elements $F$ and $D$.
The second moments are consistent with lattice QCD predictions.
\vspace{1cm}


\pagebreak


\pagebreak


\begin{table}[h]
\footnotesize
\begin{center}
\begin{tabular}{||c|r|r|r|r||}
\hline \hline
$x$    & $A_{1p}^{+}$\hspace{15mm} &  $A_{1p}^{-}$ \hspace{15mm} 
& $A_{1d}^{+}$ \hspace{15mm} &  $A_{1d}^{-}$\hspace{15mm}  \\
\hline  \hline
  0.005 &$ 0.053\pm  0.034\pm  0.005$ &$ 0.050\pm  0.038\pm  0.005$ &$-0.030\pm  0.039\pm  0.005$ &$ 0.032\pm  0.042\pm  0.005$  \\
  0.008 &$ 0.070\pm  0.034\pm  0.007$ &$-0.064\pm  0.038\pm  0.006$ &$-0.039\pm  0.038\pm  0.006$ &$ 0.013\pm  0.042\pm  0.005$  \\
  0.014 &$ 0.107\pm  0.030\pm  0.011$ &$ 0.036\pm  0.034\pm  0.004$ &$ 0.013\pm  0.033\pm  0.004$ &$-0.041\pm  0.036\pm  0.007$  \\
  0.025 &$ 0.049\pm  0.041\pm  0.005$ &$ 0.089\pm  0.048\pm  0.009$ &$-0.033\pm  0.044\pm  0.006$ &$-0.046\pm  0.049\pm  0.006$  \\
  0.035 &$ 0.125\pm  0.050\pm  0.011$ &$ 0.121\pm  0.059\pm  0.011$ &$-0.016\pm  0.053\pm  0.004$ &$ 0.020\pm  0.060\pm  0.005$  \\
  0.049 &$ 0.124\pm  0.042\pm  0.010$ &$ 0.042\pm  0.052\pm  0.005$ &$ 0.019\pm  0.046\pm  0.005$ &$ 0.142\pm  0.053\pm  0.012$  \\
  0.077 &$ 0.195\pm  0.041\pm  0.016$ &$ 0.109\pm  0.054\pm  0.009$ &$ 0.119\pm  0.046\pm  0.010$ &$ 0.076\pm  0.055\pm  0.008$  \\
  0.122 &$ 0.386\pm  0.055\pm  0.027$ &$ 0.244\pm  0.075\pm  0.018$ &$ 0.155\pm  0.063\pm  0.012$ &$ 0.045\pm  0.077\pm  0.007$  \\
  0.173 &$ 0.306\pm  0.078\pm  0.020$ &$ 0.361\pm  0.112\pm  0.027$ &$ 0.149\pm  0.093\pm  0.013$ &$ 0.138\pm  0.116\pm  0.014$  \\
  0.242 &$ 0.454\pm  0.083\pm  0.030$ &$ 0.405\pm  0.126\pm  0.032$ &$ 0.345\pm  0.104\pm  0.023$ &$ 0.183\pm  0.133\pm  0.016$  \\
  0.342 &$ 0.567\pm  0.144\pm  0.035$ &$ 0.256\pm  0.238\pm  0.021$ &$ 0.269\pm  0.190\pm  0.017$ &$-0.046\pm  0.248\pm  0.009$  \\
  0.480 &$-0.106\pm  0.245\pm  0.011$ &$ 0.438\pm  0.433\pm  0.040$ &$ 0.578\pm  0.320\pm  0.032$ &$ 1.176\pm  0.451\pm  0.063$  \\
\hline \hline
\end{tabular}
\end{center}
\caption{\em Values of semi-inclusive spin asymmetries $A_{1\,p}^+$, $A_{1\,p}^-$,
$A_{1\,d}^+$ and $A_{1\,d}^-$. The first error is statistical, the second one systematic.
 \label{asym_tab}}
\end{table}

\begin{table}[h]
\footnotesize
\begin{center}
\begin{tabular}{||c|c|c|c||c|c|c||}
\hline
\hline
  $x$   &  \multicolumn{6}{|c||}{Correlations} \\
       &  ($A_{1p},A_{1p}^+$) & ($A_{1p},A_{1p}^-$) & ($A_{1p}^+,A_{1p}^-$) &
($A_{1d},A_{1d}^+$) & ($A_{1d},A_{1d}^-$) & ($A_{1d}^+,A_{1d}^-$) \\
\hline \hline
  0.005 &  0.32 &  0.29 &  0.05 &  0.31 &  0.29 &  0.05 \\
  0.008 &  0.35 &  0.32 &  0.06 &  0.34 &  0.31 &  0.05 \\
  0.014 &  0.38 &  0.34 &  0.06 &  0.37 &  0.34 &  0.06 \\
  0.025 &  0.41 &  0.36 &  0.07 &  0.40 &  0.36 &  0.06 \\
  0.035 &  0.44 &  0.37 &  0.07 &  0.42 &  0.37 &  0.07 \\
  0.049 &  0.45 &  0.37 &  0.07 &  0.43 &  0.38 &  0.07 \\
  0.077 &  0.47 &  0.36 &  0.07 &  0.44 &  0.37 &  0.07 \\
  0.122 &  0.45 &  0.32 &  0.06 &  0.43 &  0.35 &  0.06 \\
  0.173 &  0.40 &  0.27 &  0.05 &  0.39 &  0.30 &  0.06 \\
  0.242 &  0.34 &  0.22 &  0.05 &  0.33 &  0.25 &  0.05 \\
  0.342 &  0.30 &  0.18 &  0.04 &  0.27 &  0.21 &  0.05 \\
  0.480 &  0.30 &  0.17 &  0.05 &  0.25 &  0.18 &  0.05 \\
\hline
\hline
\end{tabular}
\end{center}
\caption{\em The correlation coefficients between the asymmetries.
In the correlations only the dominating statistical uncertainties
are taken into account. Therefore the proton and deuteron asymmetries
are uncorrelated. The correlations are given by 
\mbox{${\rm cor}(A^+,A^-)=<n^+ n^->/\sqrt{<{n^+}^2><{n^-}^2>}$}
and \mbox{${\rm cor}(A,A^{+(-)})= <n^{+(-)}>/$}
\mbox{$\sqrt{<{n^{+(-)}}^2>}$},
where $n^{+(-)}$ is the number of positive (negative)
hadrons per scattered muon.
\label{cor_tab}}
\end{table}

\begin{table}[h]
\footnotesize
\begin{center}
\begin{tabular}{||c|r|r|r|c|c||} \hline \hline
$x$ &$x \Delta u_v(x)$ \hspace{14mm}   &
     $x \Delta d_v(x)$ \hspace{14mm}    &
     $x \Delta \bar q(x)$\hspace{14mm} & $\chi^2/n$ & $p$ \\
\hline \hline
\hline
  0.005 &$-0.029\pm  0.064\pm  0.017$ &$-0.105\pm  0.088\pm  0.022$ &$ 0.027\pm  0.027\pm  0.007$& 2.4/3 &0.50  \\
  0.008 &$ 0.107\pm  0.059\pm  0.023$ &$-0.050\pm  0.079\pm  0.023$ &$-0.019\pm  0.025\pm  0.009$& 4.8/3 &0.19  \\
  0.014 &$ 0.115\pm  0.049\pm  0.023$ &$-0.035\pm  0.063\pm  0.025$ &$-0.025\pm  0.021\pm  0.009$& 4.9/3 &0.18  \\
  0.025 &$ 0.004\pm  0.063\pm  0.007$ &$-0.136\pm  0.082\pm  0.019$ &$ 0.026\pm  0.027\pm  0.005$& 2.6/3 &0.46  \\
  0.035 &$ 0.017\pm  0.075\pm  0.026$ &$-0.170\pm  0.097\pm  0.035$ &$ 0.028\pm  0.032\pm  0.012$& 2.5/3 &0.47  \\
  0.049 &$ 0.029\pm  0.064\pm  0.005$ &$-0.053\pm  0.082\pm  0.011$ &$ 0.025\pm  0.027\pm  0.004$& 4.2/3 &0.24  \\
  0.077 &$ 0.171\pm  0.061\pm  0.021$ &$-0.154\pm  0.079\pm  0.025$ &$ 0.002\pm  0.026\pm  0.008$& 9.2/3 &0.03  \\
  0.122 &$ 0.287\pm  0.076\pm  0.040$ &$-0.079\pm  0.097\pm  0.045$ &$-0.026\pm  0.032\pm  0.017$& 5.1/3 &0.16  \\
  0.173 &$ 0.133\pm  0.105\pm  0.007$ &$-0.101\pm  0.132\pm  0.020$ &$ 0.035\pm  0.043\pm  0.003$& 0.6/3 &0.90  \\
  0.242 &$ 0.296\pm  0.103\pm  0.032$ &$ 0.038\pm  0.127\pm  0.039$ &$-0.021\pm  0.043\pm  0.013$& 2.1/3 &0.55  \\
  0.342 &$ 0.332\pm  0.041\pm  0.025$ &$-0.198\pm  0.078\pm  0.025$ &                            & 1.8/4 &0.77  \\
  0.480 &$ 0.205\pm  0.033\pm  0.011$ &$-0.064\pm  0.062\pm  0.010$ &                            &16.2/4 &0.003  \\
\hline
\end{tabular}
\caption{\em Values of the polarised quark distributions $x\Delta u_v(x)$, 
$x \Delta d_v(x)$ and $x \Delta \bar{q}(x)$. The distributions were obtained 
with the assumption $\Delta \bar u(x)=\Delta \bar d(x)$. 
The first errors are statistical and the second ones are systematic. 
The values in the last two bins of $x$ 
correspond to the open circles in Fig.~\ref{dq}.
In the last two columns the value of $\chi^2/n$ and the corresponding
probability $p$ are given.
\label{dq_tab}}
\end{center}
\end{table}

\begin{table}[h]
\footnotesize
\begin{center}
\begin{tabular}{||c|c|c|c||}
\hline
\hline
Error Source              &$\int_{0.003}^{0.7} \Delta u_v(x) {\rm d}x
$&$\int_{0.003}^{0.7} \Delta d_v(x) {\rm d}x$
&$\int_{0.003}^{0.7} \Delta \bar q(x) {\rm d}x$\\
\hline \hline
Beam polarisation                &   0.02        &    0.01       &   0.000           \\
Target polarisation              &   0.02        &    0.03       &   0.002           \\ 
Dilution factor                  &   0.02        &    0.01       &   0.002           \\
Acceptance variation             &   0.01        &    0.03       &   0.002           \\
Statistical error of f.f.        &   0.01        &    0.01       &   0.005           \\
Assumptions on f.f.              &   0.05        &    0.06       &   0.024           \\
Unpol. qurark dist.              &   0.02        &    0.02       &   0.001           \\
$\Delta \bar q(x) =0$            &   0.01        &    0.01       &   0.003           \\
\hline
Total systematic error           &   0.07        &    0.08       &   0.03
 \\
\hline
\hline
 Statistical error               &   0.10        &    0.14       &   0.04           \\  
\hline \hline
\end{tabular}
\caption{\em Contributions to the  error on
the integral $\int_{0.003}^{0.7} \Delta q(x) {\rm d}x$.
The abbreviation {\rm f.f.} stands for fragmentation functions.
The values of the integrals are given in Tab.\protect{\ref{int}}.
\label{syserr}}
\end{center}
\end{table}
\begin{table}[h]
\begin{center}
\begin{tabular}{||c|r|r|r||} \hline \hline
\multicolumn{4}{||c||}{Contributions to the first moments} \\
\multicolumn{4}{||c||}{of the polarised quark distributions} \\
\hline \hline
\multicolumn{4}{||c||}{} \\
\multicolumn{4}{||c||}{$\Delta \bar u(x) = \Delta \bar d(x)$} \\
\hline
     $x$        & $0 - 0.003$   & $0.003 -  0.7 \hspace{8mm}$  & $0 -  1\hspace{1cm}$
\\
\hline
$\Delta u_v$
          &   $0.04\pm0.04 $    & $ 0.73\pm0.10\pm 0.07$   &    $0.77\pm0.10\pm0.08$
    \\
$\Delta d_v$
          &  $-0.05 \pm 0.05$   &$-0.47\pm0.14\pm 0.08$    &   $-0.52\pm0.14\pm0.09$
  \\
\hline
     $x$  & $0  -  0.003$ & $0.003  -  0.3 \hspace{8mm}$ & $0  -  1\hspace{1cm}$ \\ \hline
$\Delta \bar q$
          &  $0.00\pm0.02 $   & $0.01\pm0.04\pm 0.03$   &  $ 0.01 \pm 0.04\pm0.03$
\\
\hline \hline
\multicolumn{4}{||c||}{} \\
\multicolumn{4}{||c||}{$\Delta \bar u(x) \ne \Delta \bar d(x)$} \\
\hline
     $x$        & $0 - 0.003$     & $0.003 -  0.7 \hspace{8mm}$  & $0 -  1\hspace{1cm}$  \\
\hline
  $\Delta u_v$  &  $0.04 \pm 0.04$  & $0.72 \pm 0.11 \pm 0.06$  &  $0.76 \pm 0.11 \pm 0.07 $ \\
  $\Delta d_v$  &  $-0.05 \pm 0.05$  & $-0.45 \pm 0.30 \pm 0.25$  &  $-0.50 \pm 0.30 \pm 0.25 $ \\
\hline
     $x$  & $0  -  0.003$ & $0.003  -  0.3 \hspace{8mm}$ & $0  -  1\hspace{1cm}$ \\
\hline
  $\Delta \bar u$  &  $0.00 \pm 0.02$  & $0.01 \pm 0.05 \pm 0.01$  &  $0.01 \pm 0.05 \pm 0.02 $ \\
  $\Delta \bar d$  &  $0.00 \pm 0.02$  & $0.01 \pm 0.14 \pm 0.12$  &  $0.01 \pm 0.14 \pm 0.12 $ \\
\hline \hline
\end{tabular}
\caption{\em The values for the extrapolations to low $x$ (first column),
the integrals over the measured range (second column)
and the first moments (third column).
In the last two columns the first error is statistical, the second systematic.
The upper part of the table shows the results obtained assuming
$\Delta \bar u(x) = \Delta \bar d(x)$, the lower part shows
the results without this assumption.
\label{int}}
\end{center}
\end{table}
 
\pagebreak

\end{document}

%% file: authors.tex
\pagestyle{empty} 
 
\begin {Authlist}
 		
B.~Adeva\Iref{a19},
T.~Akdogan\Iref{aa1},
E.~Arik\Iref{aa1},
A.~Arvidson\IAref{a22}{a},
B.~Badelek\IIref{a22}{a24}, 
G.~Bardin\IAref{a18}{\dagger},	
G.~Baum\Iref{a1},
P.~Berglund\Iref{a7},
L.~Betev\Iref{a12},
R.~Birsa\Iref{a21}
% ...
\end{Authlist}
\Instfoot {a1} {University of Bielefeld, Physics Department,
                33501 Bielefeld, Germany\Aref{aaa} }
\Instfoot {aa1} {Bogazi\c{c}i University and Istanbul Technical University,
		 Istanbul, Turkey\Aref{bbb} }
\Instfoot {aa111} {University of Bochum, Physics Department,
                44780 Bochum, Germany\Aref{aaa} }
\Instfoot {a2} {University of California, Department of Physics,
                Los Angeles, 90024~CA, USA\Aref{ccc}}
\Instfoot {a4} {CERN, 1211 Geneva 23, Switzerland}
\Instfoot {a5} {University of Freiburg, Physics Department,
                79104 Freiburg, Germany\Aref{aaa}}
\Instfoot {a6} {GKSS, 21494 Geesthacht, Germany\Aref{aaa}}
\Instfoot {a7} {Helsinki University of Technology, Low Temperature
                Laboratory and Institute of Particle Physics Technology,
                Espoo, Finland}
\Instfoot {a8} {University of Houston, Department of Physics,
                Houston, 77204-5506 TX, USA\AAref{ccc}{ddd}}
\Instfoot {a9} {JINR, Dubna, RU-141980 Dubna, Russia} 
\Instfoot {a10} {University of Mainz, Institute for Nuclear Physics,
                 55099 Mainz, Germany\Aref{aaa}}
\Instfoot {a11} {University of Mons, Faculty of Science,
                 7000 Mons, Belgium}
\Instfoot {a12} {University of Munich, Physics Department,
                 80799 Munich, Germany\Aref{aaa}}
\Instfoot {a13} {Nagoya University, CIRSE and Department of Physics, Furo-Cho,
                 Chikusa-Ku, 464 Nagoya, Japan\Aref{eee}}
\Instfoot {a14} {NIKHEF, Delft University of Technology, FOM and Free University,
                 1009 AJ Amsterdam, The Netherlands\Aref{f{}f{}f}}
\Instfoot {a16} {Northwestern University, Department of Physics,
                 Evanston, 60208 IL, USA\AAref{ccc}{ddd}}
\Instfoot {a18} {C.E.A.~Saclay, DAPNIA, 91191 Gif-sur-Yvette, France\Aref{ggg}}
\Instfoot {a19} {University of Santiago, Department of Particle Physics,
                 15706 Santiago de Compostela, Spain\Aref{hhh}}
\Instfoot {a20} {Tel Aviv University, School of Physics,
                 69978 Tel Aviv, Israel\Aref{iii}}
\Instfoot {a21} {INFN Trieste and
                 University of Trieste, Department of Physics,
                 34127 Trieste, Italy}
\Instfoot {a22} {Uppsala University, Department of Radiation Sciences,
                 75121 Uppsala, Sweden}
\Instfoot {a23} {University of Virginia, Department of Physics,
                 Charlottesville, 22901 VA, USA\Aref{ccc}}
\Instfoot {a24} {Soltan Institute for Nuclear Studies
                 and Warsaw University,
                 00681 Warsaw, Poland\Aref{jjj}}
\Instfoot {a26} {Yale University, Department of Physics,
                 New Haven, 06511 CT, USA\Aref{ccc}}
\Anotfoot {a} {Now at Gammadata, Uppsala, Sweden}
\Anotfoot {b} {Now at University of Virginia, Department of Physics,
                 Charlottesville, 22901 VA, USA\Aref{ccc}}
\Anotfoot {c} {Now at CIEMAT, Avda Complutense 22, 28040, Madrid, Spain}
\Anotfoot {d} {Now at Fermi National Accelerator Laboratory,
               Batavia, 60510 IL, USA}
\Anotfoot {e} {Now at University of Wisconsin, USA}
\Anotfoot {f} {Now at NIKHEF P.O.B. 41882, 1009 DB Amsterdam, The Netherlands}
\Anotfoot {g} {Now at SLAC, Stanford 94309 CA USA}
\Anotfoot {h} {Permanent address: Miyazaki University, Faculty of Engineering,
               889-21 Miyazaki-Shi, Japan}
\Anotfoot {i} {Permanent address: Paul Scherrer Institut, 5232 Villigen, 	
		   Switzerland}
\Anotfoot {j} {Permanent address: The Institute of Physical and 
               Chemical Research (RIKEN), wako 351-01, Japan}
\Anotfoot {k} {Permanent address: University of California,
                 Institute of Particle Physics,
                 Santa Cruz, 95064 CA, USA}
\Anotfoot {l} {Permanent address: KEK, Tsukuba-Shi, 305 Ibaraki-Ken,Japan}
\Anotfoot {m} {Now at University of Michigan, Ann Arbor MI48109, USA}
\Anotfoot {n} {Now at CERN, 1211 Geneva 23, Switzerland}
\Anotfoot {o} {Permanent address: Northeastern University, Department of
               Physics, Boston, 02115 MA, USA} 
\Anotfoot {p} {Now at Penn. State University, 303 Osmond Lab,
               University Park, 16802 PA, USA}
\Anotfoot {q} {Permanent address: Rice University, Bonner Laboratory, 
                Houston, TX 77251-1892, USA}
\Anotfoot {r} {Permanent address: University of Buenos Aires,
               Physics Department, 1428 Buenos Ai\-res, Argentina }
\Anotfoot {s} {Permanent address: Warsaw University of Technology, Warsaw,
               Poland}
\Anotfoot {t} {Now at F.Hoffmann-La Roche Ltd., CH-4070 Basel, Switzerland}
\Anotfoot {u} {Now at Los Alamos National Laboratory, Los Alamos, NM 87545, USA}
\Anotfoot {aaa} {Supported by the Bundesministerium f\"ur Bildung,
               Wissenschaft, Forschung und
               Technologie}
\Anotfoot {bbb} {Partially supported by TUBITAK and the Centre for 
               Turkish-Balkan Physics Research and Application 
               (Bogazi\c{c}i University)}
\Anotfoot {ccc}  {Supported by the U.S. Department of Energy}
\Anotfoot {ddd}  {Supported by the U.S. National Science Foundation}
\Anotfoot {eee}  {Supported by  Monbusho Grant-in-Aid
                for Scientific Research (International Scientific Research
                Program and Specially Promoted Research)}
\Anotfoot {f{}f{}f} {Supported by the National Science Foundation (NWO)
               of The Netherlands}
\Anotfoot {ggg} {Supported by the Commissariat \`a l'Energie Atomique}
\Anotfoot {hhh} {Supported by Comision Interministerial de Ciencia
               y Tecnologia}
\Anotfoot {iii} {Supported by the Israel Science Foundation.}
\Anotfoot {jjj} {Supported by KBN SPUB/P3/209/94 and /P3/21/97}
\Anotfoot {\dagger} {Deceased.}